\documentclass[final,5p,times,twocolumn]{elsarticle}
\usepackage{graphics}

\journal{Atroparticle Physics}
\begin{document}

\begin{frontmatter}

\title{CdWO$_4$ scintillating bolometer for Double Beta Decay: Light and Heat anticorrelation,
light yield and quenching factors}

\author[INFN]{C. Arnaboldi}
\author[LBL]{J.W. Beeman}
\author[INFN]{O. Cremonesi}
\author[INFN,UNIMIB]{L. Gironi}
\author[INFN,UNIMIB]{M. Pavan}
\author[INFN]{G. Pessina}
\author[INFN]{S. Pirro \corref{cor1}}\ead{Stefano.Pirro@mib.infn.it}
\author[INFN]{E. Previtali}

\cortext[cor1]{Corresponding author}
\address[INFN]{INFN - Sezione di Milano Bicocca I 20126 Milano - Italy}
\address[UNIMIB]{Dipartimento di Fisica - Universit\`{a} di Milano Bicocca I 20126 Milano - Italy}
\address[LBL]{Lawrence Berkeley National Laboratory , Berkeley, California 94720, USA}

\date{\today}

\begin{abstract}

We report the performances of a 0.51 kg CdWO$_4$ scintillating bolometer to be used for future Double Beta Decay Experiments.
The simultaneous  read-out of the heat and the scintillation light allows to discriminate between different interacting 
particles aiming at  the disentanglement and the reduction of background contribution, key issue for next generation experiments.
We will describe the observed  anticorrelation between the heat and the light signal and we will show how this feature can be used in 
order to increase the energy resolution of the bolometer over the entire energy spectrum, improving up to  a factor 2.6 on the 2615 keV 
line of $^{208}$Tl. The detector was tested in a 433 h background measurement that permitted to  estimate extremely low internal trace 
contaminations of $^{232}$Th and $^{238}$U. The light yield of  $\gamma/\beta$, $\alpha$'s   and neutrons is  presented.
Furthermore we developed a method in order to  correctly evaluate  the absolute thermal quenching factor of $\alpha$-particles in scintillating 
bolometers.
 
\end{abstract}


\end{frontmatter}

\section{Introduction}

Double Beta Decay (DBD) searches became of critical importance after the discovery of the neutrino oscillations and plenty of experiments 
are now in the construction phase and many others are in R\&D  phase \cite{DBDgeneral1,DBDgeneral2,DBDgeneral3,DBDgeneral4}. 
The main challenges for all the different experimental techniques are the 
same \cite{EPJA-2006}: i) increase the active mass, ii) decrease the background, and iii)  increase the energy resolution.

Thermal bolometers are  ideal detectors for this survey: crystals  can be grown with different  interesting DBD-emitters and, fundamental 
for next generation experiments, they show an excellent  energy resolution. 

The CUORICINO Experiment \cite{PRC-2008}, constituted by an array of 62 TeO$_2$ crystal bolometers, demonstrated not only the power of 
this technique but also that the main source of background for these detectors arises from surface contaminations of radioactive 
$\alpha$-emitters. Moreover  simulations show that this contribution will largely dominate the expected background of  the CUORE 
Experiment \cite{Arnaboldi2004,fondoBB} in the region of interest, since there is no possibility to separate this background from the two DBD electrons. 
The natural way to discriminate this background is to use a scintillating bolometer \cite{PHAN-2006}. 
In such a  device the simultaneous and independent read out of the heat and the scintillation light permits to discriminate events 
due to $\gamma$/$\beta$,
$\alpha$ and neutrons  thanks to  their different scintillation yield.
Moreover  if the crystal is based on a DBD emitter whose transition energy exceeds the 2615 keV $\gamma$-line of $^{208}$Tl then the 
environmental background due to natural $\gamma$'s will decrease abruptly.

CdWO$_4$ is an ideal candidate for such kind of detector:
\begin{itemize}
\item it is a well established scintillator
\item $^{116}$Cd (7.5 \% i.a) has a DBD transition at 2805 keV
\item the light yield (LY) is rather large
\item the radiopurity of this compound is ``naturally'' high
\end{itemize}

Due to these favourable features this crystal compound was already used to perform a DBD experiment \cite{Danevich-2003} using 
``standard'' Photomultipliers.  A large mass  experiment, based on enriched CdWO$_4$ crystals, readout by Photomultipliers, was also 
proposed  \cite{Bellini2000}. But this technique,  limited by the modest achievable energy resolution, key point of future 
experiments, is no more pursued.

\section{Experimental details}
\label{sec:Experimental-details}
The dimensions of the  CdWO$_4$ crystal tested  are 4 cm diameter, 5 cm height. All the surfaces of the crystal  are polished at optical grade.
The crystal was tested as a standard scintillator  at room temperature. It was wrapped with a reflecting sheet (3M Radiant Mirror film VM2000)
and coupled with optical grease to a Photomultiplier (Hamamatzu R6233). The  energy resolution evaluated on the $^{137}$Cs line is 12.3 \% FWHM.

\begin{figure}
\begin{center}
\resizebox{0.35\textwidth}{!}
{\includegraphics{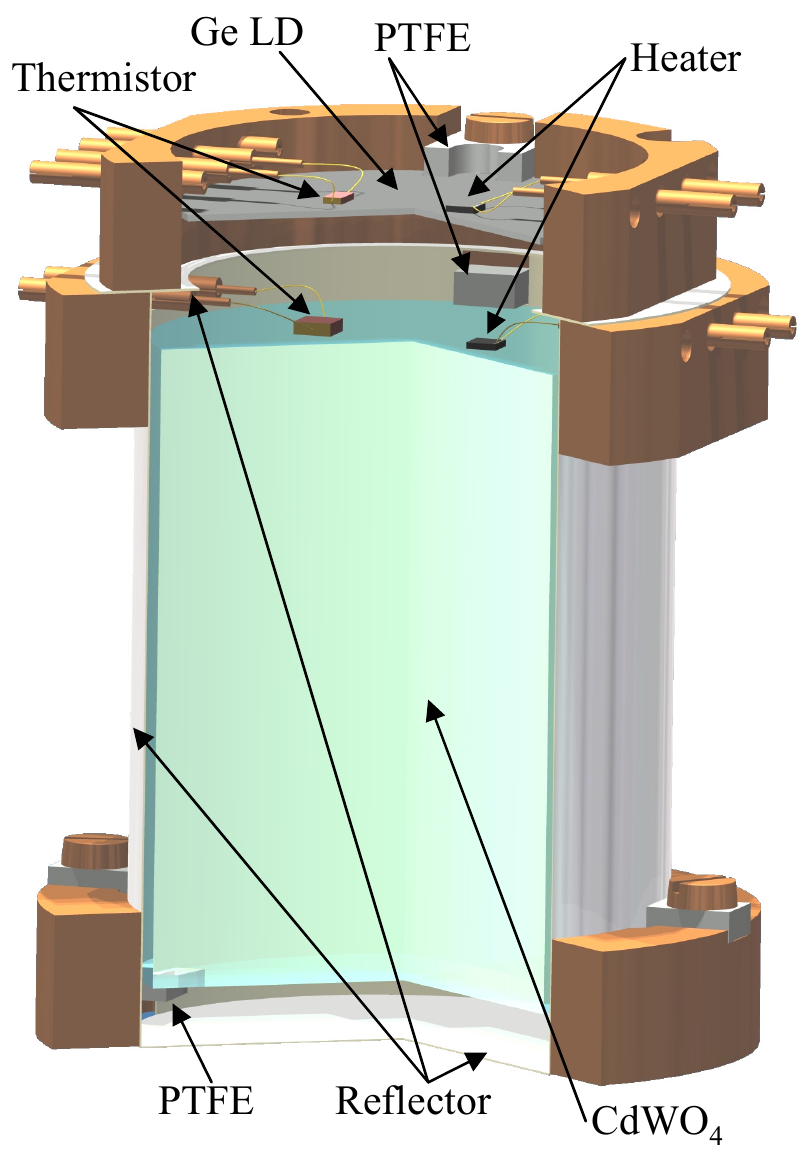}}
\caption{Setup of the detector. The two Cu columns, holding the two frames, are not visible from the chosen perspective.}
\label{fig:fig1}      
\end{center}
\end{figure}

Our detector setup is schematized in Fig.~\ref{fig:fig1}. 
It  is held by means of four L-shaped Teflon pieces fixed to the two cylindrical Cu frames. 
The frames are held together through two Cu columns. 
The crystal is surrounded (without being in thermal contact) by a  43 mm diameter reflecting sheet (3M  VM2002).
The Light Detector (LD) \cite{NIMA-2006}  is constituted by a  36 mm diameter, 1 mm thick pure Ge crystal absorber. 
The surface of the Ge facing the crystal is further ``darkened'' through the deposition of a 600 \AA\ layer of SiO$_2$ in order 
to increase the light abortion.
Furthermore since the diameter of the  LD is smaller with respect to the one  of the reflecting cavity, a 
``reflecting ring'' has been mounted in order to decrease light losses. On the opposite face of the crystal a reflecting sheet is mounted.
The temperature sensor of the CdWO$_4$ crystal  is a 3x3x1 mm$^3$ neutron transmutation doped Germanium, the same used 
in the CUORICINO experiment.
The temperature sensor of the LD has smaller volume (3x1.5x0.4 mm$^3$) in order to decrease its heat capacity, increasing therefore its 
thermal signal.
A resistor of  $\sim$300 k$\Omega$, realized with a heavily doped  meander on a 3.5 mm$^3$ silicon chip, is attached to  
each absorber and acts as a heater to  stabilize the gain of the bolometer \cite{ALES98,Arna2003}. 
The detectors were operated deep underground in the Gran Sasso National Laboratories in the CUORE R\&D test cryostat. 
The details of the electronics and the cryogenic facility can be found elsewhere \cite{NIMA-2006-B,NIMA-2006-C,NIMA-2004}.

The heat and light pulses, produced by a particle interacting in  the CdWO$_4$ crystal and transduced in a voltage pulse by 
the NTD thermistors, are amplified and fed into a 16 bit NI 6225 USB ADC unit. 
The entire waveform of each triggered voltage pulse is sampled and acquired.
The time window has a width of 256 ms  sampled with 512 points. The trigger of the CdWO$_4$ is software generated while 
the LD is automatically acquired in coincidence with  the former. 
The amplitude and the shape of the voltage pulse is then determined by the off line analysis that makes use of the 
Optimal Filter technique.
The energy calibration of the CdWO$_4$ crystal is performed  using $\gamma$ sources placed outside the cryostat. 
The Heat axis is calibrated  attributing to each identified $\gamma$
peak the nominal energy of the line, as if all the energy is converted into heat. Consequently this
calibration does not provide an absolute evaluation of the heat deposited in the crystal.
The dependency of amplitude from energy is parameterized with a second order polynomial in log(V)
where V is the heat pulse amplitude. The three coefficients of the polynomial are fitted on calibration
data (the heat pulse amplitude corresponding to the known $\gamma$ lines visible in the spectrum). The
choice of such a function was established by means of simulation studies based on a thermal model of
the detectors.
\begin{figure}
\begin{center}
\resizebox{0.48\textwidth}{!}
{\includegraphics{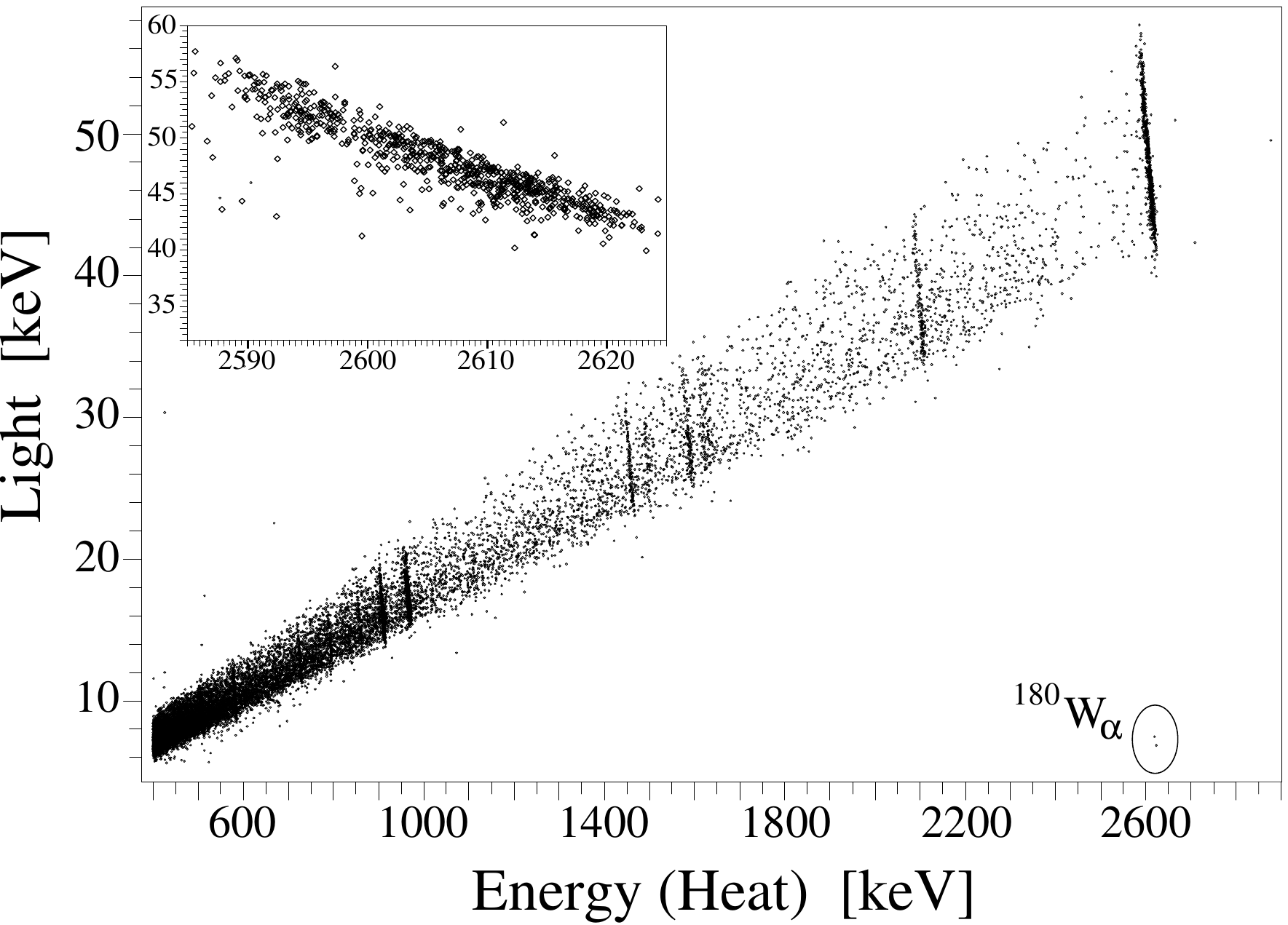}}
\caption{Scatter plot Light vs. Heat obtained in a 96 h calibration using an external $^{232}$Th source. In the inset the 
highlight of the $^{208}$Tl line. In the circle two events due to the internal $\alpha$-decay of  $^{180}$W. The energy 
spectrum below 400 keV is completely dominated by the  $^{113}$Cd 
$\beta$-decay, with a rate close to 0.4 Hz.}
\label{fig:fig2}      
\end{center}
\end{figure}

The energy calibration of the LD is obtained thanks to a  weak $^{55}$Fe source placed close to the Ge that illuminates 
homogeneously the face opposed to the  CdWO$_4$ crystal. During the LD calibration its trigger is set independent from 
the one of the CdWO$_4$.  The LD  is calibrated using a simple linear function.  
The FWHM energy resolution of the LD, evaluated on the X doublet at  5.90 and 6.49  keV, is 480 eV FWHM.
In the range of 0$\div$50 keV, i.e. the energy interval  of the light signals,  the  response  of the LD can be definitely 
assumed to be linear.  

Three sets of data have been collected with this device: two calibrations using  $^{232}$Th and  $^{40}$K sources, a long 
background measurement (433 h) and a neutron measurement (8 h) done exposing the detector to an Am-Be source.

\section{Light-Heat scatter plot}
\label{sec:Light-Heat scatter plot}
\begin{figure}
\begin{center}
\resizebox{0.48\textwidth}{!}
{\includegraphics{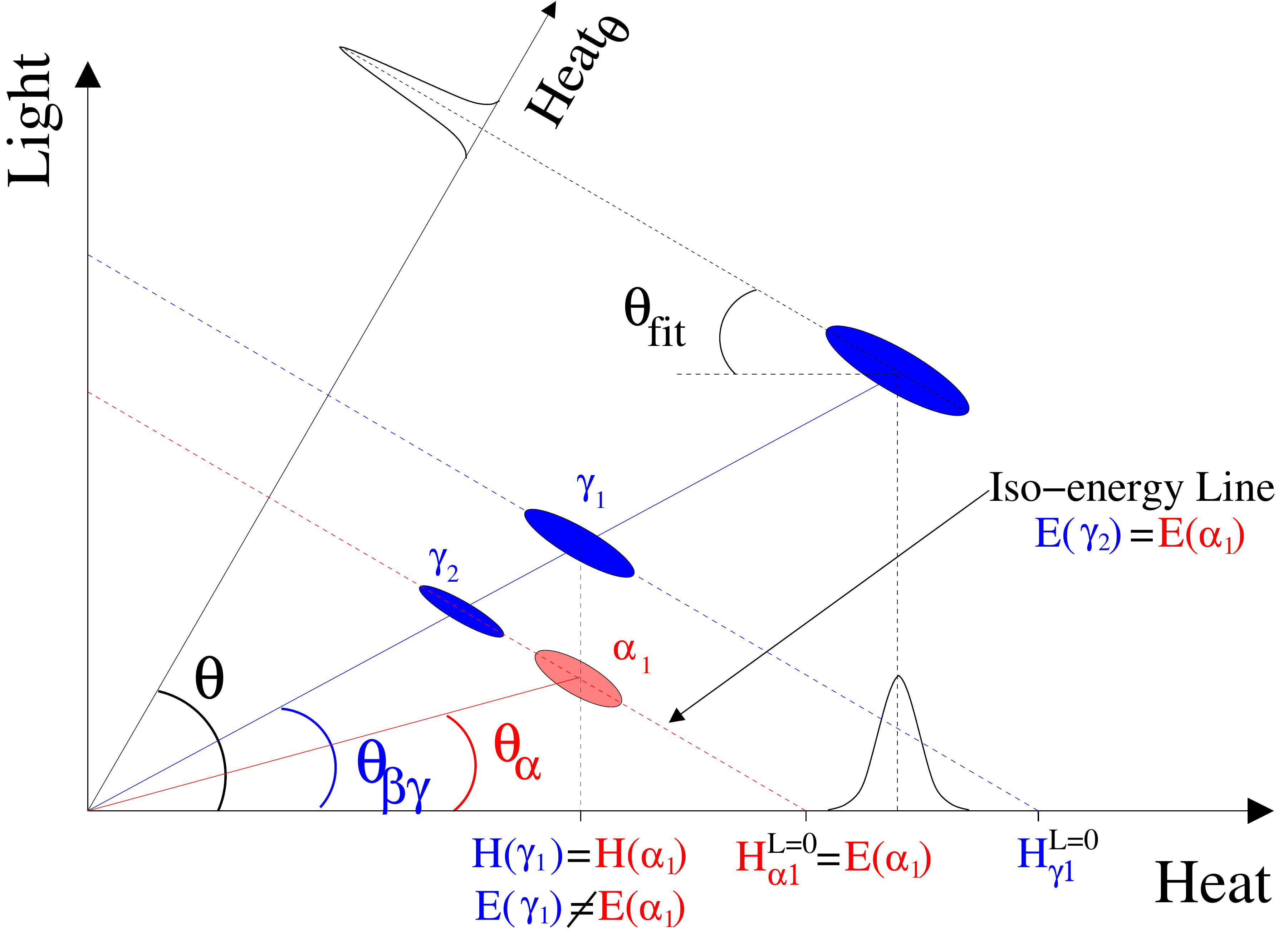}}
\caption{Light vs. Heat scheme for the interpretation of the energy correction method. The blue (dark) spots represent the $\gamma/\beta$ monochromatic
events, while the red (light) represents an $\alpha$.}
\label{fig:fig3}      
\end{center}
\end{figure}

In this field the usual way to present the results is to draw the Light vs. Heat scatter plot. 
Here each event is identified by a point with abscissa equal to the heat signal (recorded by the CdWO$_4$ bolometer), and 
ordinate equal to the light signal (contemporary recorded by the LD). In the scatter plot, $\gamma/\beta$, $\alpha$ and 
neutrons give rise to  separate bands, in virtue of their characteristic Light to Heat ratio  (see Figs.~\ref{fig:fig5} 
and \ref{fig:fig6}). This feature is the result of the different LY's characterizing these particles: $\gamma/\beta$ events belong to a 
distribution (see Fig.~\ref{fig:fig3}) characterized by the $\theta_{\beta\gamma}$ angle, while the $\alpha$ events are characterized by  $\theta_{\alpha}$.

The scintillation Quenching Factor (QF) is defined as the ratio of the scintillating yield of an  interacting 
particle  ($\alpha$, neutron, nucleus) with respect to the LY of a $\gamma$/$\beta$ event at the same energy.
From Fig.~\ref{fig:fig3} we have QF$_\alpha$=tan($\theta_{\alpha}$)/tan($\theta_{\beta\gamma}$). But this holds only in first approximation, as will be exposed
in Sec.~\ref{sec:Light Yield and Quenching Factors} and Sec.~\ref{sec:Heat absolute scale and Heat Quenching Factor}.

Within each band, monochromatic events (i.e. those corresponding to the complete absorption of a monochromatic particle 
in the CdWO$_4$ crystal) appear as sections of lines with negative slope, showing a strong anticorrelation 
between Heat and Light. This is evident in the Light vs. Heat scatter plot obtained with a source of $^{232}$Th (Fig.~\ref{fig:fig2}) 
where different  $\gamma$-lines are clearly visible. Despite the much lower statistics,  the same feature is perceptible also for the $\alpha$ 
lines that appear in the background spectrum (Fig.~\ref{fig:fig7}).

A simple model accounts for the observed pattern. The energy E of a monochromatic particle, fully absorbed in the CdWO$_4$ crystal, is divided 
into two channels: a fraction ($L$) is spent in the production of Light (photons) and a fraction ($H$) is spent in the production of Heat (phonons). 
In absence of ``blind'' channels (i.e. channels in which the energy deposited into the crystal is stored in some system that do not takes part
in signal formation) the energy conservation 
requires:
\begin{equation}
\label{eq:energy-split}
E=E_{Heat}+E_{Light}= (1-k)E+kE=H+L
\end{equation} 
being $k$  the fraction of the total energy that escapes the crystal in form of light. 
The last equation states not only the energy conservation, but also a subtle distinction 
between the Heat (what is measured) and the energy definition.

Monochromatic events should  produce a spot in the scatter plot, with a size determined by the intrinsic energy resolution of the heat and light detectors.
The observed spread along a negative slope line, however,  can be accounted for assuming the existence of fluctuations in the H and L signal amplitudes. 
Since the energy has to be conserved the amplitude of the fluctuations ($\delta$L and $\delta$H) must compensate each other: $\delta$L=-$\delta$H. 
In other words the fluctuations are anti-correlated between each other and produce the observed pattern. They are visible because 
their amplitude is much larger than the intrinsic resolution of the detectors.

We can devise two mechanisms that introduce a fluctuation in the L/H ratio:
i) the statistical Poissonian fluctuation in the emitted light: the more the energy spent in light
production, the less the energy spent in heat;
ii) any variation of the LY as those due to position dependent effects (inhomogeneities and
defects of the crystals that modify the light emission or self absorption of scintillation photons).
Regardless the mechanism, we will demonstrate, in the next section, how this anticorrelation can be used in order to increase (correct) the 
energy resolution of the detector over the entire energy spectrum.

Fig.~\ref{fig:fig3} tries to summarize the model here discussed. The dark spots represent monochromatic energy deposition in the CdWO$_4$ 
crystal, as those observed during a $\gamma$ calibration. For each of them a negative slope (iso-energy) line is tracked. 
The intercept on the Heat axis (H$^{Light=0}$) is the Heat that would correspond to a full heat conversion of the deposited energy 
(in absence of light emission or in the case in which all the light is absorbed by the crystal itself).
In this simple model  $\gamma/\beta$'s and  $\alpha$'s  releasing the same energy  within the CdWO$_4$ crystal lie on the same iso-energy line. 
But due to the larger LY, $\gamma/\beta$ events will convert less heat within the crystal with respect to an $\alpha$ particle 
releasing the same energy: part of the energy escapes the crystal in form of photons.
This feature can be easily deduced from Fig.~\ref{fig:fig3}: the $\alpha_{1}$ particle release the same total energy of the $\gamma_{2}$ 
particle (they belong to the same iso-energy line) but they show a different position on the Heat axis. In fact the $\alpha_{1}$ shows the same heat of 
the $\gamma_{1}$ particle that releases a larger energy within (but not ``into'') the crystal.
The ``usual'' picture (in which $\gamma/\beta$ and  $\alpha$ of the same energy show the same heat signal) corresponds to the limit in which the light emitted 
is negligible (or completely re-absorbed by the crystal): in this case the distinction between H and E becomes meaningless.

It is important to remark that up to now  it was   assumed  to be able to measure the \emph{absolute} values of energy converted into heat and light.

Actually, the measured experimental heat and light signals can be written as 
\begin{equation}
\label{eq:light_and_heat}
\begin{array}{ll}
H=\alpha(1-k)E \\
L=\beta \epsilon kE 
\end{array}
\end{equation} 
being  $\alpha$ and $\beta$ the  absolute calibration factors for the Heat and Light axis. The factor $\epsilon$, instead, takes into account the 
overall light collection efficiency.

Under the condition that $\alpha$ and $\beta$ are not depending on the energy, and considering that the conservation of energy
requires that $\delta[kE]$=- $\delta[(1-k)E]$, Eq. ~\ref{eq:light_and_heat} implies that

\begin{equation}
\label{eq:tanteta}
\frac{\delta L}{\delta H}= - \frac{\beta\epsilon}{\alpha} = - tan(\theta_{fit})
\end{equation}

The last equation states that the effect of  miss-calibrations of the Heat and Light axis  simply implies a variation of the slope of monochromatic lines, 
which is exactly what we observe experimentally. Also the light collection efficiency plays a fundamental role in the evaluation of the slope, as discussed
in the next section.

It is important to note that, unlike the Heat channel, the LD energy calibration is more delicate.
The  energy calibration on this detector is performed using \emph{ionizing} X-Rays from the  $^{55}$Fe source. \emph{A priori} the thermal signal 
that arises by the absorption of 1 keV of photons at the boundaries of the Ge crystal (moreover covered by the SiO$_2$ layer) could 
give a different thermal signal with respect to a 1 keV ionizing energy. 
In any case this effect still preserves proportionality, so that it can be considered within the factor $\beta$.

\section{Light-Heat anticorrelation: energy resolution}

\label{sec:Light-Heat anticorrelation} 
The energy anticorrelation between the two signals of a double readout system was already observed 
in ionization/scintillation detectors \cite{Conti-2003} as well as in heat/scintillation bolometers \cite{APL-2005} with a device 
very similar to the one presented here. In the former case this anticorrelation was demonstrated to improve the energy resolution
by a factor close to $\sim$ 25 \% (evaluated @570 keV). In  paper \cite{APL-2005}
two different $\gamma$-lines were studied at rather low energy. An evident anticorrelation was found but, however, the correlation
factor of the two lines was found to strongly depend on the energy.

The energy correction procedure is graphically  explained in  Fig.~\ref{fig:fig3}. 
Here blue (dark) and red (light) spots represent $\gamma/\beta$ and $\alpha$ produced by monochromatic events. In order to obtain a
spectrum with improved energy resolution we have to combine the heat  and light values  of each single event in an appropriate way.
This can be obtained in a natural way performing a simple SU(1) rotation in the scatter plane an projecting the points  on the 
Heat$_{\theta}$ axis:

\begin{equation}
\label{SU(1)}
\begin{array}{ll}
$Heat$_{\theta}= \;\;\;$Heat cos$\theta \;  + \; $Light sin$ \theta		 \\
$Ligh$_{\theta}=-$Heat sin$\theta \; + \; $Light cos$ \theta
\end{array}
\end{equation}

The value of $\theta$ can be evaluated in two different ways: i) it can be ``optimized'' in order to obtain the best energy resolution 
on the Heat$_{\theta}$ axis (energy minimization); ii) by fitting the monochromatic spots in the scatter plot with a negative 
slope line, evaluating  the $\theta_{fit}$ angle for each single distribution (for construction we have $\theta=  \pi/2 -  \theta_{fit}$)

It is clear from this scheme that some  assumptions  should be satisfied  in order for this procedure to work properly:
\begin{itemize}
\item the slope of each monochromatic  spot (i.e. the $\theta_{fit}$  angle) or the $\theta$ angle minimizing
       the FWHM has to be the same within all the lines;
\item the LY of the different class of interacting particles ($\alpha$'s and $\gamma/\beta$) has to be independent from energy.
\end{itemize}

The first condition ensures that the rotation minimizes the FWHM on the \emph{whole} energy spectrum.
It has to be noted that if the first condition is not verified, this technique still provides a useful tool for resolution 
optimization. In case of different $\theta_{fit}$ values  the rotation angle will be chosen in order to reach the best performances in the region of 
more interest for the physics (e.g. the region where the DBD peak should appear). 

The second condition ensures that the projected spectrum keeps linearity and energy calibration.
This point is less trivial and will be treated in more detail in Sec.~\ref{sec:Light Yield and Quenching Factors}.
\begin{figure}
\begin{center}
\resizebox{0.48\textwidth}{!}
{\includegraphics{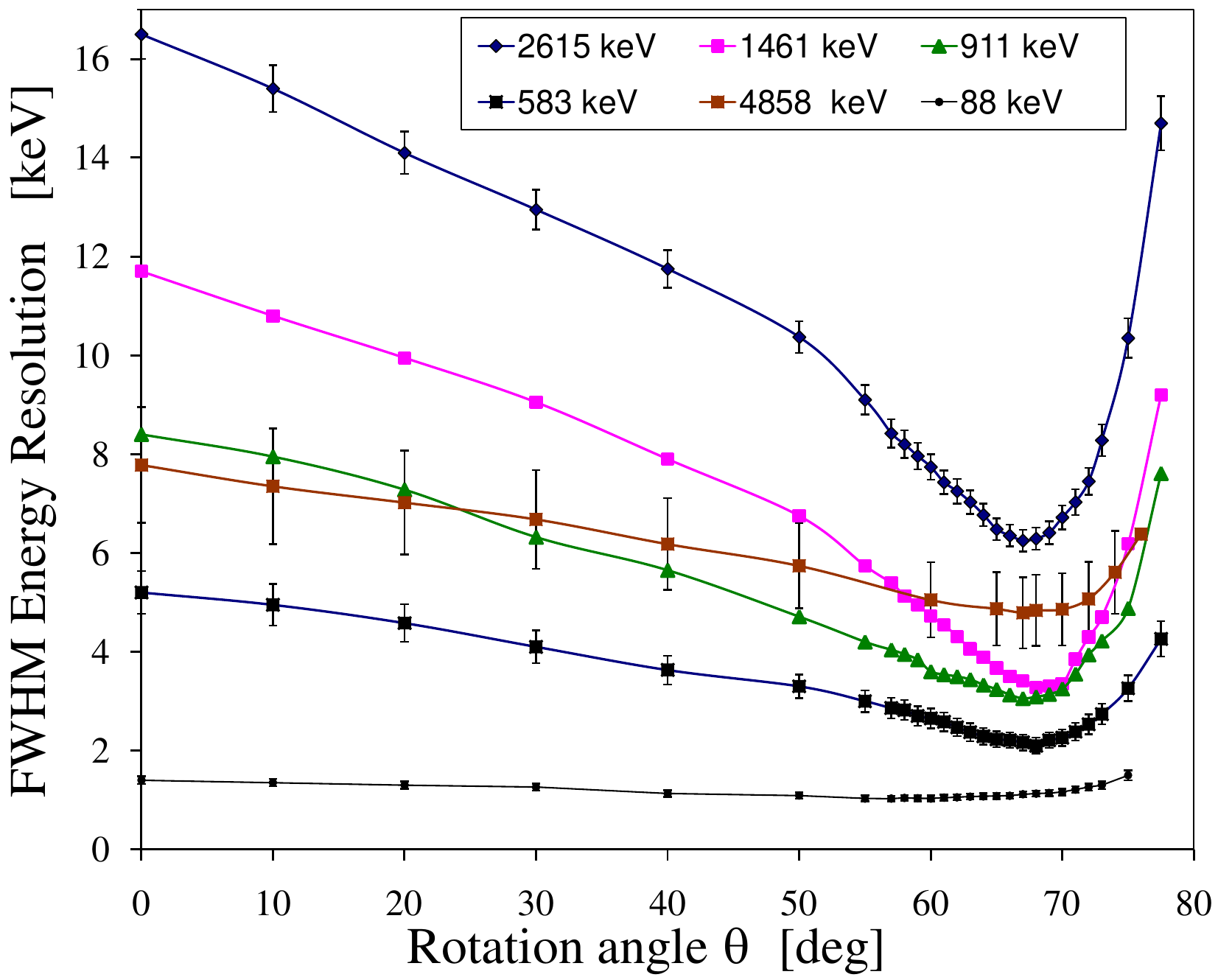}}
 \caption{FWHM energy resolutions evaluated for different $\theta$ angles. For each angle a spectrum is produced and the
corresponding peaks are fitted. The curves through the data points are only to guide eyes. For two peaks the corresponding
error on the FWHM is not plotted for better clarity.}
\label{fig:fig4}      
\end{center}
\end{figure}

We applied both the methods in order to evaluate the rotation angle.
Using the energy minimization method, the value of the rotation $\theta$ angle is allowed to vary over a wide range (0-80 degrees) and for each
angle a spectrum is produced projecting the data on the Heat$_{\theta}$ axis. Finally the FWHM energy resolution of the more
intense peaks is evaluated by a fit (we use an asymmetric Gaussian reporting then the average FWHM).

We applied this method not only to the calibration spectrum but also to the   background measurement.
In this way we were  able to study also the  the 88 keV  $\gamma$-line of $^{109}$Cd (accidentally contamination of our 
CdWO$_4$ sample) and  the  $\alpha$ peaks due to   internal trace contaminations: in this latter case we observe a 
line at  the full energy of the decay  (Q=$\alpha$+nuclear recoil).
The  internal $\alpha$-lines observed are $^{234}$U-$\alpha$(Q=4858 keV), $^{180}$W-$\alpha$ (Q=2516 keV) 
and  $^{238}$U-$\alpha$ (Q=4270 keV).

In Fig.~\ref{fig:fig4} we report the FWHM energy resolutions evaluated on the most intense lines obtained varying 
the rotation angle $\theta$ and projecting on Heat$_{\theta}$ . The  energy resolutions of the peaks are evaluated for each 
single rotated spectrum. The minimum of the energy resolution occurs within  $\theta$=(67.5 $\pm$ 0.5) deg for all the 
lines, except for the 88 keV Electron Capture internal line of  $^{109}$Cd.

In fact the minimum of the energy resolution for this line, 1.03 $\pm$ 0.06 keV FWHM  is obtained at  $\theta$= 60 deg
while at 67.5 deg it becomes 1.11 $\pm$ 0.06 keV. These two values seem rather similar, but from Fig.~\ref{fig:fig4} it turns out that 
the resolution curve for this peak  shows a different behaviour with respect to all the other lines. We do not have an explanation
for this enhancement in the anticorrelation at low energies, but  the same effect was already observed in \cite{APL-2005}.   

Each monochromatic events distribution  in the scatter plot was also fitted with a linear function, as previously discussed.

The $\theta_{fit}$ angles as well as the  most relevant parameters of this analysis are summarized in Tab.~\ref{Tab:tab1}.
The values of the $\theta_{fit}$ angles for the $\alpha$'s and $\gamma$/$\beta$'s lines reported here are compatible within the
experimental errors within the two groups (with the exception of the 88 keV line). However, with respect to the $\gamma$/$\beta$'s,  the 
angular coefficient calculated by fitting linearly the single lines gives a  result that is systematically slightly larger with respect to 
the value (67.5$\pm$0.5)  obtained through the energy minimization. 
This systematic slightly difference could arise from the fact that while the bidimensional fit is model independent, the energy minimization depends on the 
fitting procedure of the peaks.  
Moreover it should be pointed out that the quoted error on $\theta_{fit}$  don't include  systematic effects (introduced, for example, by the 
continuum background) whose values could dominate the total error. 

The values of the  $ \theta _{fit}$ angles reported in Tab.~\ref{Tab:tab1} seems to show  a slightly difference between  $\alpha$ and $\gamma/\beta$.
However it has to be  pointed  out  that the quoted error on the $\theta_{fit}$ for the $\alpha$ particles has a (further) systematic error that cannot 
be easily evaluated. The point is that the  $\alpha$ events were collected during the long background measurement. During this period the CdWO$_4$
crystal was slowly cooling down (the LD, on the contrary, was rather stable during the same period). This drift, as explained in 
Sec.~\ref{sec:Experimental-details},  is usually corrected through the Heater pulse. 
If the temperature drift is rather small ($\lesssim$ 50  $\mu$K) then the correction is independent from the energy.
On the contrary, if the drift is larger (in this case the drift was $\approx$ 0.3 mK ) then the shape of the signal changes appreciably. In such cases 
the correction starts to depend on the energy. As a consequence a small drift in function of the time (or the temperature) becomes appreciable. 
Moreover this drift depends on the energy. So,  in order to evaluate the second, third and last column in Tab.~\ref{Tab:tab1} each 
$\alpha$ peak was ``self-stabilized''. This can introduce an uncontrolled systematics since we are somehow forcing a distribution to obtain a
minimum. On the other hand, a check was made on   the weak 2615 keV $\gamma$-line present in the background spectrum. As in the case of the $\alpha$'s, 
we had to self-correct the drift. The obtained $\theta _{fit}$ value is  71.5$\pm$0.4, rather consistent with the $\alpha$ background.
For the following we will assume $\theta _{fit}^{\gamma/\beta} = \theta _{fit}^{\alpha}$.
The $^{232}$Th and $^{40}$K calibrations,  as well as the 88 keV analysis, on the other hand, were made at the end of the measurement 
in a rather stable working temperature.

As a final remark it has to be noticed that  the projection of  the events on the Heat$_{\theta}$  axis  introduces an Heat (energy) normalization factor $K$ that will be 
different for different classes of  particles, namely K$_{\gamma\beta,\alpha}$=cos$\theta$(1+tan$\theta$ tan$\theta_{\gamma\beta,\alpha}$).
This feature   is rather evident in Fig.~\ref{fig:fig3}: in the Heat axis we have H($\alpha 1$)=H($\gamma 1$)
while in the rotated Heat$_\theta$ axis we have H($\alpha 1$)=H($\gamma 2$).
This has two consequences: the first, obvious, is that the intercalibration between  $\alpha$ and $\gamma/\beta$ is different within 
the Heat spectrum and the Heat$_{\theta}$ spectrum. The second is that the same consideration holds within the 
same class of events: if the LY of $\alpha$ particles (for example)  depends slightly on energy, then the linearity 
of the projected spectrum will change.

\section{Light Yield and Scintillation Quenching Factor}
\label{sec:Light Yield and Quenching Factors}

In Tab.~\ref{Tab:tab1} we also present the values of the LY, determined on the different monochromatic lines.  
We define the LY as the energy released in the LD (in keV) for a nominal  energy deposition of 1 MeV in the scintillating crystal.

\begin{table}[htb]
\begin{center}
\caption{Table with the main parameters of the detector. The theoretical FWHM energy resolution of the CdWO$_4$ crystal,
evaluated through Optimal Filtering, is 0.8 keV.  The LY of the $\alpha$-particles (last three lines) is evaluated for 
the  \emph{overall} Q-value of the decay. The  correct value (assuming negligible scintillation for the nucleus recoil) 
is $\approx$ 2\% larger. \dag See text.}
\label{Tab:tab1}
\begin{tabular}{@{}lcccc}
\hline
Line & FWHM$_{\theta=0}$     & FWHM$_{\theta=67.5}$    & LY 				& $\pi/2-\theta _{fit}$  \\

[keV] &     [keV] 			 &      [keV] 			   & [keV/MeV]			& [deg]			         \\

\hline
88     &  1.4$\pm$0.08   	 & 1.11$\pm$0.06           &  17.3$\pm$0.2      &  61.6$\pm$0.9          \\
583    &  5.2$\pm$0.4		 & 2.1$\pm$0.2             &  17.0$\pm$0.35     &  70.8$\pm$0.4          \\
911    &  8.4$\pm$0.5   	 & 3.05$\pm$0.2            &  17.5$\pm$0.2      &  70.6$\pm$0.3          \\
1461   &  11.7$\pm$0.5       & 3.27$\pm$0.2            &  17.45$\pm$0.15    &  70.4$\pm$0.3          \\ 
2615   &  16.5$\pm$0.5       & 6.25$\pm$0.22           &  17.6$\pm$0.1      &  69.8$\pm$0.2          \\
2516   &  5.2$\pm$1.6  \dag  & 4.3$\pm$1.3 \dag        &  2.74$\pm$0.08   	&  73.0$\pm$0.2 \dag     \\
4270   &  7.0$\pm$1.5  \dag  & 4.4$\pm$1.0 \dag 	   &  3.29$\pm$0.07     &  72.3$\pm$0.2 \dag     \\
4858   &  7.8$\pm$1.2  \dag  & 4.8$\pm$0.7 \dag        &  3.44$\pm$0.08     &  72.7$\pm$0.2 \dag     \\
\hline
\end{tabular}
\end{center}
\end{table}

The table shows that the LY of the $\gamma/\beta$ events is, within the error, the same for all the observed lines (including the 
88 keV line that shows a different resolution curve in Fig.~\ref{fig:fig4}). On the other hand a small, but rather clear, energy dependence is 
evident for  $\alpha$ particles. This effect can also be observed in Fig.~\ref{fig:fig6} in which it is evident that the $\alpha$ particles
belong to a curve rather than a straight line.  
Moreover, as already stated, the observed $\alpha$-lines are internal so that the effect 
cannot be ascribed to surface effects.
A theoretical explanation for this behaviour and a detailed discussion of the energy dependence of LY and QF in different scintillating 
crystals can be found in \cite{Tretyak2010}. 
Essentially, the observed behaviour reflects the fact that an $\alpha$ particle scintillates less with respect to an electron because of its 
larger dE/dx which can induce saturation effects in the scintillator (the Birks law \cite{Birks}). The energy dependence of the LY and QF is a 
consequence of the energy dependence of the stopping power. Electrons, due to their low stopping power, do not suffer of saturation effects and 
their LY is, consequently, energy independent. For this reason the definition of QF$_\alpha$=tan($\theta_{\alpha}$)/tan($\theta_{\beta\gamma}$), given in 
Sec.~\ref{sec:Light-Heat scatter plot}, represents only an approximation.

Using the values in the table  we can evaluate  the scintillation Quenching Factors of $\alpha$ particles (QF=LY($\alpha$)/LY($\beta$)):
QF$_{^{180}W}$=0.160$\pm$.006, QF$_{^{238}U}$=0.192$\pm$.006, QF$_{^{234}U}$=0.201$\pm$.006. 

Within this framework, furthermore, it is possible to evaluate the total amount of light $k$ that escapes the crystal.

Using  Eq.~\ref{eq:tanteta} and considering that, for definition, LY$\equiv$L/E=$\beta \epsilon k$, we obtain:

\begin{equation}
\label{eq:absLY1}
LY=tan(\theta_{fit}) \alpha k
\end{equation}

Now we have to note that the  energy calibration we adopted for the Heat axis (as discussed in Sec.~\ref{sec:Experimental-details}) implies
H$\equiv$E. This means (see Eq.~\ref{eq:light_and_heat}) that $\alpha$=1/(1-k$_{\gamma}$), being k$_{\gamma}$ the absolute LY for $\gamma/\beta$ events.
Using this last relation in the last equation we finally get for  $\gamma/\beta$

\begin{equation}
\label{eq:absLY}
k_\gamma=\frac{LY_\gamma}{tan(\theta_{fit})+LY_\gamma}
\end{equation}

Using the values of Tab.~\ref{Tab:tab1} we get k$_\gamma$= 4.6 \%

As a final remark, we point out that the large spread in the light channel (6.8 \%  at the 2615 keV scintillation signal), as observed in the 
scatter plots, cannot be  dominated by the fluctuation of  the Poissonian statistics of the absorbed photons. In fact we have that the mean energy absorbed 
by the LD  for a 2615 keV interaction in  CdWO$_4$ is of the order of 2.615$\cdot$17.6= 46 keV. Assuming $\approx$ 3 eV/photon we get (assuming a Fano 
factor=1)  the variation in the light channel to be of the order of 1/$\sqrt{46000/3}$=0.8 \%. 
Moreover in other tests performed with different CdWO$_4$ crystal samples we observed the same anticorrelation (in terms of $\theta_{fit}$) but with a much 
smaller spread. Thus the  magnitude of this spread is probably dominated by  fluctuation in the overall light collection efficiency. 
In particular we believe that the grade of the surfaces of the crystal (in terms of diffusion of the scintillation light) plays 
an important role in this mechanism.

\section{Heat absolute scale and Heat Quenching Factor}
\label{sec:Heat absolute scale and Heat Quenching Factor}
From the previous sections is evident that for scintillating bolometers the Heat/Energy scale is different for different kinds of 
interacting particles (or, better, for particles characterized  by different LY's). 
In other words, with our convention for the Heat axis calibration, H measures 
the total energy of a $\gamma/\beta$ particle while it doesn't for an $\alpha$: the energy calibration of $\alpha$-particles has to be dealt in a separate way.

This ``displacement'' for the $\alpha$  lines is evident in our experimental data.
In particular the $^{180}$W-$\alpha$ line (as can be seen in Fig.~\ref{fig:fig2}), whose nominal Q-value is 2516~keV \cite{Cozzini2004}, appears 
(in the $\gamma$-calibrated spectra) at 2627$\pm 2$ keV.
This line is rather peculiar since it definitely proves that the mis-calibration cannot be ascribed to a wrong extrapolation of the detector 
amplitude vs. energy calibration (the highest $\gamma$-line used for the calibration of the Heat signal is indeed the $^{208}$Tl line at 2615 keV).

The observed shift in keV,  $\Delta E=E_{\alpha}-E_{\gamma/\beta}$, between the experimentally reconstructed energy and its nominal value is given - for 
the three mentioned $\alpha$'s - by:

\begin{equation}
\label{eq:energy-shift}
\Delta E_{^{180}W}=110\pm3  \; \Delta E_{^{238}U}=172\pm3.5   \; \Delta E_{^{234}U}=180\pm2   \;\;\;\;
\end{equation}

This shift can be easily recognized by looking at Fig.~\ref{fig:fig3}. Let us consider the case in which the crystal does not scintillate: in this case the two 
particles, $\alpha$1 and $\gamma$1 will show, as example, $E(\alpha 1)  <  E(\gamma 1 )$. If  the crystal scintillates and  the light  
escapes the crystal, then the two events will move left on the two different iso-energy lines up to reach the case in which H($\alpha$1)=H($\gamma$1). This 
means that the energy shift can be written as  $\Delta E=H^{L=0}_{\gamma 1} - H^{L=0}_{\alpha 1}$. By simple geometric consideration we get

\begin{equation}
\label{eq:energy-gamma}
 E^{Light=0}_{\gamma/\beta}=E+\frac{Light_{\gamma/\beta}}{ tan(\theta_{fit})}=E \left(1+\frac{LY_{\gamma/\beta}}{ tan(\theta_{fit})}\right)
\end{equation}

So, finally, we obtain

\begin{equation}
\label{eq:DeltaE}
\Delta E =  E \;[MeV] \; \cdot  \left(\frac{LY_{\gamma/\beta}-LY_\alpha} {tan(\theta_{fit})}\right)  \;  \; \; [keV]
\end{equation}

Inserting the experimental values we get:

\begin{equation}
\label{eq:energy-shift-rec}
\Delta E_{^{180}W}=107\pm 3  \; \Delta E_{^{238}U}=175\pm 6  \; \Delta E_{^{234}U}=195\pm 7 \;\;\;\;
\end{equation}

These last values are in very good agreement with the experimental ones in Eq.~\ref{eq:energy-shift}, confirming our model.

To conclude this section, we need to face an important question about the definition of the Heat Quenching Factor (HQF). 
The conventional way to treat differences in the response of a detector with respect to different type of interacting  particles is to introduce relative 
quenching factors. This is exactly what is done for the scintillation light. In the same way, following what is often done in literature 
\cite{Ales-97, Coron-2008,Ang-2005,Benoit-2007}, it is possible to introduce the HQF  as the ratio between the Heat signal 
of an $\alpha$ particle and that of a $\gamma/\beta$ of the same energy, fully absorbed in the detector. With this definition, for our detector we can 
quote the HQF measured for the $^{180}$W $\alpha$ decay as: HQF$_{^{180}W}$=2627/2516=1.044.
It is straightforward to show  that in our framework we can easily evaluate 
HQF=1+$\frac{LY_{\gamma/\beta}-{LY_{\alpha}}}{\tan(\theta_{fit})} = 1+\frac {LY_{\gamma/\beta}}{\tan(\theta_{fit})}(1-QF_{\alpha})$.

Even if the definition of the (scintillation) QF and the HQF is identical, due to the different physical mechanism that generates them, they show a 
\emph{substantial} difference. While the former depends, substantially, only on the ``nature'' of the scintillator, the HQF, instead,  depends on the amount 
of light (energy)  that escapes the crystal: it depends on the size and/or on the quality of the crystal. It is also obvious that since some (minor) 
amount of light that escapes the crystal can be reflected by the reflecting foil and can be absorbed by the crystal itself, then the HQF will depend also by 
the setup itself.

With the  3x3x2 cm$^3$ CdWO$_4$ crystal measured in  \cite{PHAN-2006}, we found the position of the $^{180}$W line at 2670 keV, giving therefore HQF=1.061. 

Now one relevant question that could be addressed is: do we have blind channels in our detector? We don't have a general method to investigate 
this point, but the  results presented so far are still valid even if we assume that a fixed fraction of the total energy is lost in blind channels (we used only Heat/Light ratios
in our framework). 
Then we can investigate if this fraction is different between a $\gamma/\beta$ and an $\alpha$. In other words we could ask ourselves if the sum of the heat 
and light signals is the same for particle converting the same total energy in the scintillator. We provide therefore an alternative definition of HQF as the 
ratio of E$^{Light=0}_{\alpha}$ and E$^{Light=0}_{\gamma/\beta}$.

Provided that our model that compensates the energy losses due to scintillation is correct, using
the values of Eq.~\ref{eq:energy-shift-rec} we can evaluate the (corrected)  HQF:
 
HQF($^{180}$W)=1.001$\pm$ .002 HQF($^{238}$U)=1.000 $\pm$ .002 and HQF($^{234}$U)=0.997 $\pm$ .002.

These HQF's are compatible with 1 meaning that there are no intrinsic differences in the heat signal generated by an $\alpha$
with respect to a signal generated by a $\gamma/\beta$\footnote{We note here that although we always discuss of $\alpha$'s it is true that we 
are always considering the system ($\alpha$ + nuclear recoil), where however the nuclear recoil transports only a small fraction of the total energy.}. 
Alternatively we can state that any eventually present blind channel have the same behaviour for the two.
\begin{figure}
\begin{center}
\resizebox{0.48\textwidth}{!}
{\includegraphics{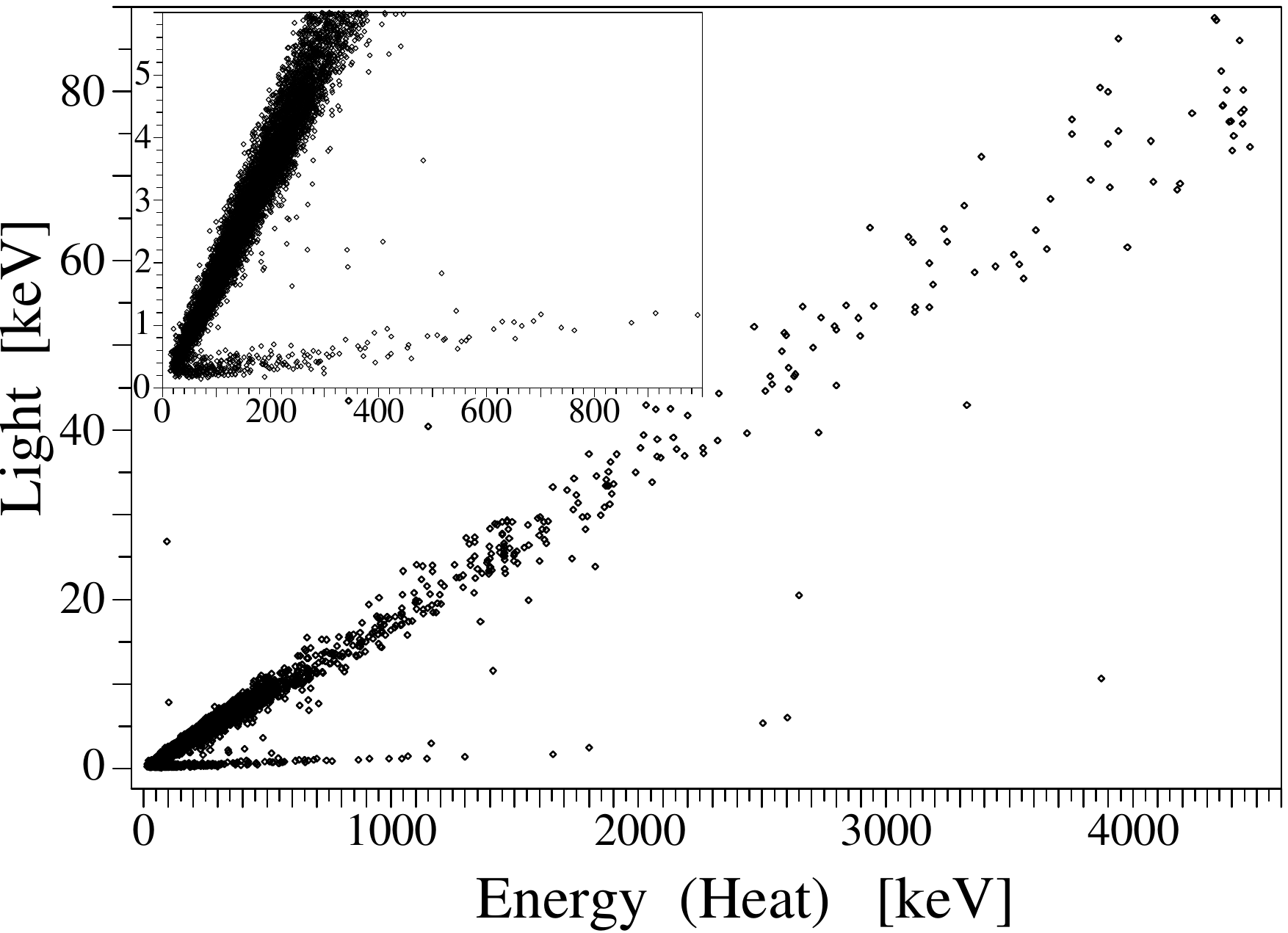}}
\caption{Scatter plot Heat vs. Light obtained in a 8 hour measurement with an Am-Be neutron source. In the inset the low energy region. The
energy spectrum extends up to 9 MeV. The 4.44 MeV ``line'' due to the $^{12}$C$^*$ is visible. The intrinsic energy spread of this 
peak is rather large ($\approx$ 1 \%) due to Doppler shift  induced by the ``in flight''  decay of $^{12}$C$^*$ in the source.}
\label{fig:fig5}      
\end{center}
\end{figure}

\section{Background Rejection: $\alpha$'s and neutrons}

The detector was also exposed to a neutron source. It consists of a 185 kBq Am-Be source with a neutron production rate of $\approx$10 n/s. 
The neutron spectrum has its maximum at $\sim$5 MeV with an high energy tail reaching 10 MeV.
The source was placed inside the shielding of the cryostat, in the same place where the $\gamma$ sources are inserted. 
The  scatter plot is shown in  Fig.~\ref{fig:fig5}:  neutron direct interactions are clearly visible, especially in the inset. 
The $\gamma$/$\beta $ events extends well above  2615 keV due to  (n,$\gamma$) reaction on the surrounding  materials but, mainly,  by
the source itself: for each neutron produced there is 60 \% probability  to produce an excited state of   $^{12}$C that emits 
a $\gamma$ of 4.44 MeV.

Moreover it has to be remarked (see also Sec.~\ref{sec:Background measurement: internal contaminations}) that $^{113}$Cd has a huge neutron capture 
cross section, with a Q-value larger than 9 MeV, so  that ``mixed events'' are possible (a neutron scatters on the Oxygen of the 
crystal and then is absorbed by the $^{113}$Cd with subsequent de-excitation of the nucleus).
From Fig.~\ref{fig:fig5} we evaluate the neutron scintillation QF with respect to $\gamma$/$\beta$ to be (0.14$\pm$0.03).
The error is dominated by the systematics induced by the extreme weakness of the scintillation light in the region 0$\div$300 keV, where most of the events are 
recorded.

\begin{figure}
\begin{center}
\resizebox{0.48\textwidth}{!}
{\includegraphics{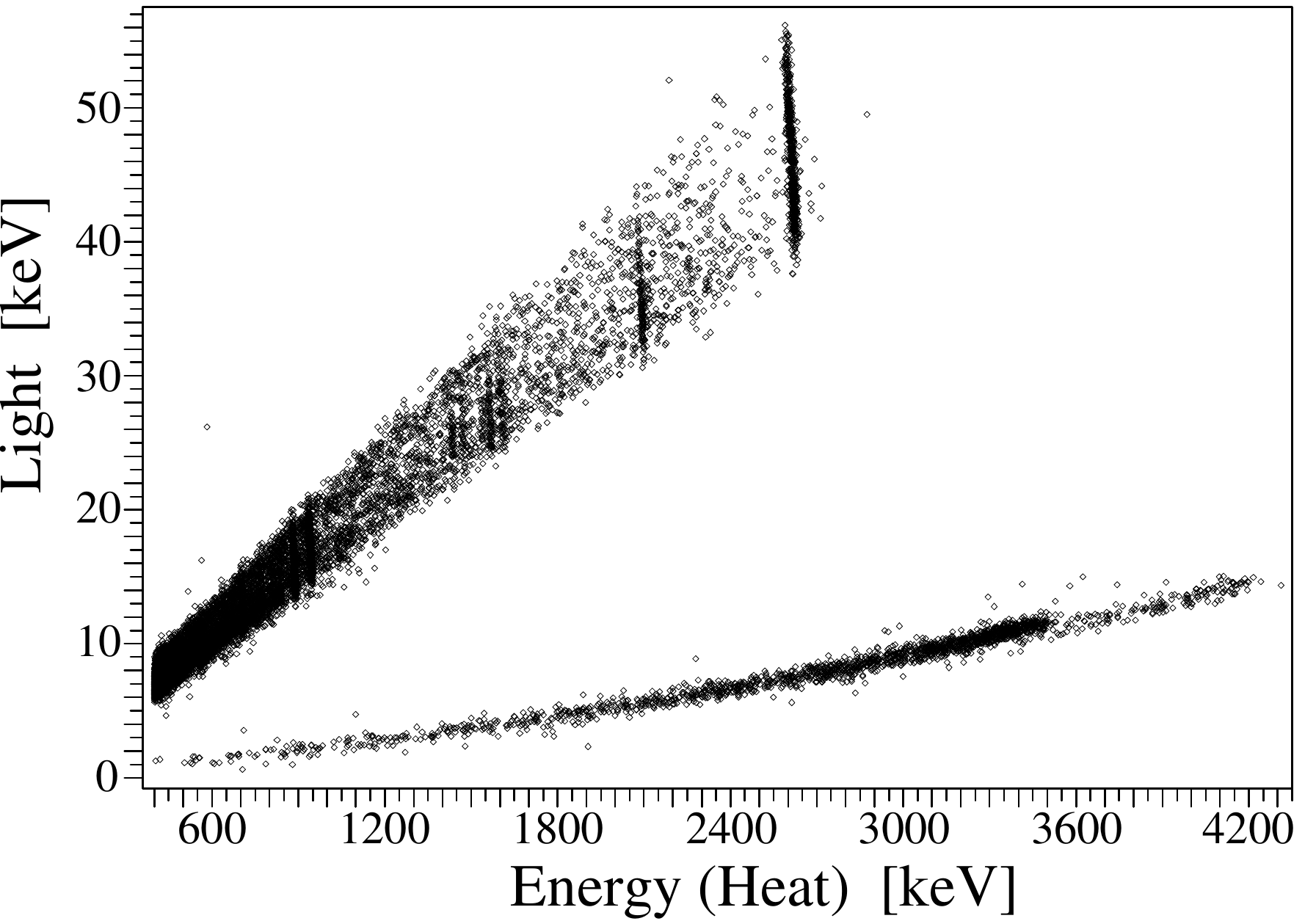}}
\caption{Scatter plot Light vs. Heat obtained in a 93 hours $^{232}$Th calibration measurement with a  smeared $\alpha$ source 
facing the CdWO$_4$. The liquid standard also contains a small amount of $^{234}$U. The alpha curve is completely 
separated from the $\gamma$/$\beta$ line. Moreover it  can be noted that the  $\alpha$ events are not belonging to a 
straight line, showing a decrease of the QF with the energy, as discussed in Sec.~\ref{sec:Light Yield and Quenching Factors}.}
\label{fig:fig6}      
\end{center}
\end{figure}

In a subsequent test, an $\alpha$ source was mounted close to the detector in order to evaluate its rejection capability.
The custom source was build up using 2 $\mu$L of a 0.1 \% $^{238}$U  liquid Standard. The  U ``droplets'' were dried on 
a 1 cm$^{2}$ Al tape and then covered with a 6 $\mu$m Aluminized Mylar foil in order to ``smear'' the alpha energy down to the 
region of interest.  The source was then faced to the crystal on the  face opposed to the LD. 
In Fig.~\ref{fig:fig6} we present the scatter plot obtained with the above mentioned source during a  93 hours $^{232}$Th 
calibration measurement. Fixing at 4$\sigma$ the acceptance on the light signal of the 2615 keV $\gamma$-line, we evaluate a 
rejection factor for $\alpha$ particles  $>$35$\sigma$ at the same energy. 
As far as  the  fast neutron interaction is concerned , the  rejection factor will  be obviously larger, but  the most dangerous contribution  will arise 
from thermal neutron absorption through $^{113}$Cd. But, at least in principle, thermal neutron can be effectively shielded while fast 
neutron, via $\mu$-spallation within the shielding or close to it, cannot.

\section{Background measurement: internal contaminations}
\label{sec:Background measurement: internal contaminations}
\begin{figure}
\begin{center}
\resizebox{0.48\textwidth}{!}
{\includegraphics{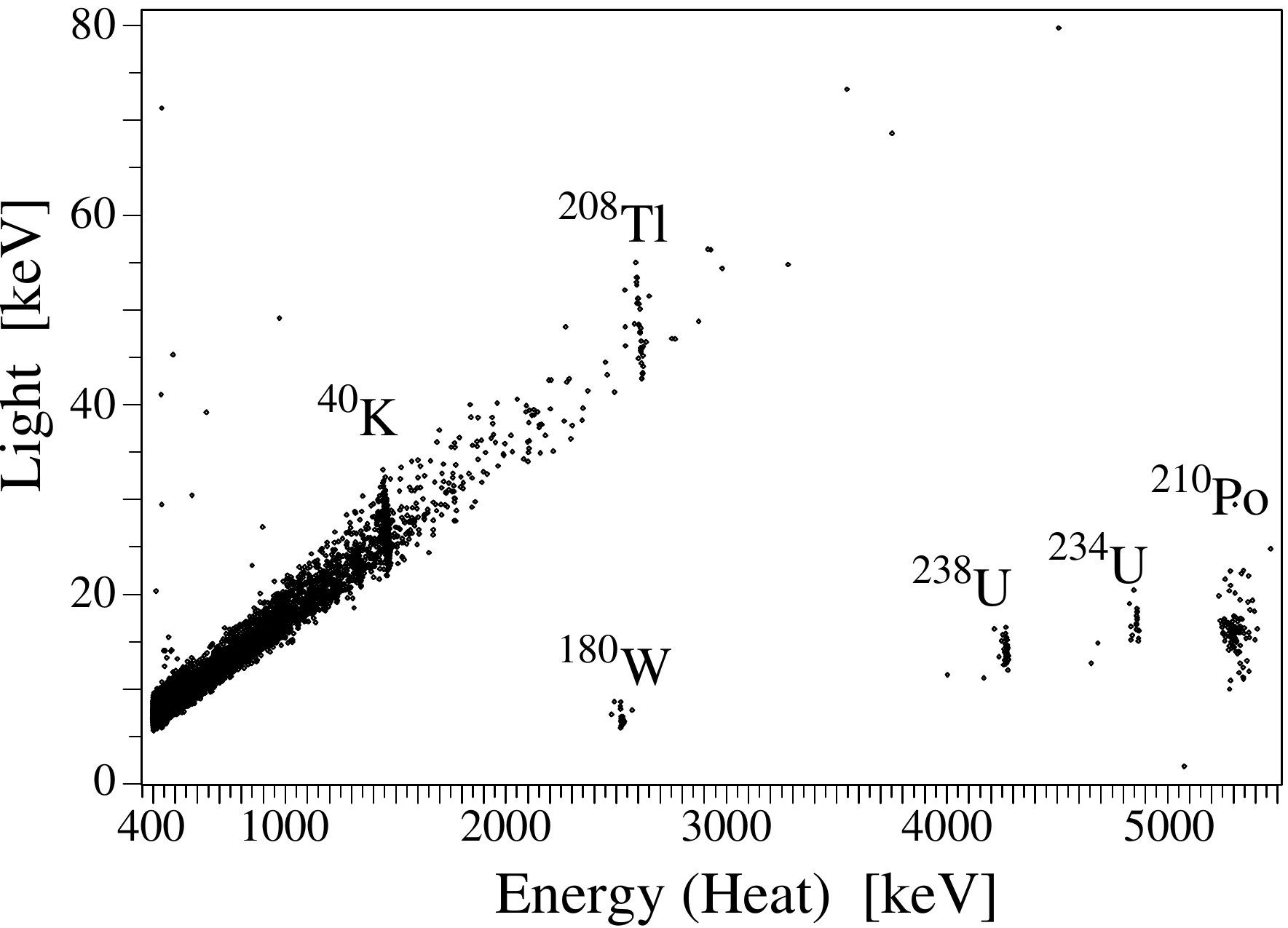}}
\caption{Scatter plot Light vs. Heat obtained in a 433 hours live time background measurement. Part of the $^{210}$Po ``bump'' is 
characterized  by smaller light emission. This could arise from a worse light collection efficiency, especially from the lateral 
part of the crystal. The energy scale of the $\alpha$ is corrected according with what exposed in  Sec.~\ref{sec:Heat absolute scale and Heat Quenching Factor}.}
\label{fig:fig7}      
\end{center}
\end{figure}

In Fig.~\ref{fig:fig7} we present the Light vs. Heat scatter plot obtained in the 433 h live time background measurement. 
It can be easily recognized that we observe only two transitions belonging to the Uranium chain, namely $^{238}$U and 
$^{234}$U, corresponding to an internal contamination of 3.1$\cdot$10$^{-12}$ g/g and   5.7$\cdot$10$^{-17}$ g/g, respectively.
The Uranium chain is broken at $^{234}$U, while a  contamination  of $^{210}$Po is clearly seen. As it often 
happens, it is rather difficult to determine if this Po contamination arise from its parent, $^{210}$Pb. Moreover the absence of
a clear peak indicates that the contamination is partially penetrating (few $\mu$m) the surface of the crystal (or its surroundings).
No events can be ascribed to the Thorium chain, giving a limit of 9$\cdot$10$^{-13}$ g/g (95\% CL) for $^{232}$Th, one of the most 
``dangerous'' contaminant for DBD searches.
But the main (unexpected) feature resulting from Fig.~\ref{fig:fig7} is the $\gamma$/$\beta$ background above  2615 keV. 
These 11 events cannot arise from external $\gamma$'s due to $^{208}$Tl or $^{214}$Bi since the observed background lines 
(583 keV, 2615 keV and 1764 keV) are too weak to allow such contribution at high energy. The same holds for internal 
contaminations belonging to Th and U chain. 
Unrecognized pile-up of the $^{113}$Cd-$\beta$ decay (Q= 320 keV) with a 2615 keV $\gamma$ cannot contribute at this 
level (we expect $\sim$ 2 ev/year). 
Internal contaminations due to rare high-energy $\beta$ emitters  like  $^{106}$Rh (Q=3541 keV) are extremely difficult to evaluate.
But, in any case, we observe  $\gamma$/$\beta$ events up to 4.5 MeV, extremely large to be associated with 
standard ``known''  $\beta$ emitters.
On the other hand we previously tested  different CdWO$_4$ scintillating crystals \cite{Gironi09}, obtaining, with larger 
statistics, no events above 2615 keV. The main difference with respect to \cite{Gironi09} is that in the present work the 
environmental neutron shielding (consisting of 7 cm of polyethylene and 1~cm of CB$_4$) was  removed from the top of the 
cryostat, leaving $\sim$0.6~m$^2$ opening, corresponding to $\approx$ 8 \% of the total neutron shielding. We think that the 
thermal neutron absorption by $^{113}$Cd is the most probable mechanism in order to explain the anomalous background above 2615 keV.

\section{Conclusions}
For the first time a ``large'' scintillating bolometer to be used for future DBD searches was fully characterized in details in terms of 
energy resolution, internal contaminations and particle identification capabilities.
It was  shown how the use of the anticorrelation between light and heat improves the energy resolution by a factor 1.26 @ 88 keV and 
up to a factor 2.6 @ 2615 keV.
We developed, for the first time, a procedure that evaluates the Heat Quenching Factor corrected for the loss of the scintillation light.  

This CdWO$_4$, grown without any precaution in terms of radiopurity,  shows extremely low trace contaminations in U and Th. 
Nonetheless a future experiment based on this compound should use isotopic enrichment, for two reasons: i) increase the  
mass of  $^{116}$Cd; ii) decrease the fraction of $^{113}$Cd in order to avoid  $\beta$ decay and, more 
important, decrease (n,$\gamma$) reaction due to $^{113}$Cd.

As a final remark we point out that such a crystal, depleted in $^{113}$Cd, would be an extremely interesting compound 
not only for DBD experiments but also  for Dark Matter searches.

\section {Acknowledgements}
The results reported here have been obtained in the framework of the Bolux R\&D Experiment funded by INFN, aiming  
at the optimization of a cryogenic DBD Experiment for a next generation experiment.
Thanks are due to E. Tatananni, A. Rotilio, A. Corsi and B. Romualdi  for continuous and constructive help in the  overall 
setup construction. Finally, we are especially grateful to Maurizio Perego for his invaluable help in the development and improvement  of the 
Data Acquisition software.


\begin{thebibliography}{00}



\bibitem {DBDgeneral1}V. I. Tretyak  and Yu. G. Zdesenko, Atomic Data and Nuclear Data Tables, \rm \bf 80 \rm (2002): 83;
\bibitem {DBDgeneral2}S. Elliott and P. Vogel, Ann. Rev. Nucl. Part. Sci. \rm \bf 52 \rm (2002):115;
\bibitem {DBDgeneral3}A. Morales and J. Morales, Nucl. Phys. B (Proc. Suppl.) \rm \bf 114 \rm (2003):141;
\bibitem {DBDgeneral4}F. T. Avignone III, S. R. Elliott and J. Engel (2006), Rev. Mod. Phys., \rm \bf 80 \rm (2008):481. 
\bibitem {EPJA-2006} S. Pirro, Eur. Phys. J. \rm \bf A 27 \rm, S1 (2006):25
\bibitem {PRC-2008} C. Arnaboldi, {\it et al.}, Phys. Rev. \rm \bf C 78 \rm (2008):035502
\bibitem {Arnaboldi2004} C. Arnaboldi, {\it et al.}, Nucl. Instr. and Meth. \rm \bf A 518 \rm(2004):775  
\bibitem {fondoBB} M. Pavan , {\it et al.}, Eur. Phys. J \rm \bf A 36 \rm (2008):159
\bibitem {PHAN-2006} S. Pirro, {\it et al.}, Physics of Atomic Nuclei \rm \bf 69  \rm No.12 (2006):2109
\bibitem {Danevich-2003} F.A. Danivich, {\it et al.}, Phys. Rev. \rm \bf C 68 \rm (2003):035501
\bibitem {Bellini2000} G. Bellini,  {\it et al.}, Phys. Lett. \rm \bf B 493 \rm (2000):216
\bibitem {NIMA-2006} S. Pirro, {\it et al.}, Nucl. Instr. and Meth. \rm \bf A 559 \rm(2006):361
\bibitem {ALES98} A. Alessandrello, {\it et al.}, Nucl. Instr. and Meth. \rm \bf A 412 \rm(1998):454
\bibitem {Arna2003} C. Arnaboldi, G. Pessina, E. Previtali, IEEE Tran. on Nucl. Sci.  \rm \bf 50 \rm(2003):979
\bibitem {NIMA-2006-B} S. Pirro, Nucl. Instr. and Meth. \rm \bf A 559 \rm(2006):672
\bibitem {NIMA-2006-C} C. Arnaboldi, G. Pessina, S. Pirro, Nucl. Instr. and Meth. \rm \bf A 559 \rm(2006):826
\bibitem {NIMA-2004} C. Arnaboldi, {\it et al.},  Nucl. Instr. and Meth. \rm \bf A 520 \rm(2004):578
\bibitem {Conti-2003} E. Conti, {\it et al.}, Phys Rev.  \rm \bf B 68\rm (2003):054201
\bibitem {APL-2005} J. Amare, {\it et al.}, Appl. Phys. Lett. \rm \bf 87 \rm (2005):264102
\bibitem {Tretyak2010} V.I. Tretyak, Astropart. Phys.   \rm \bf  33 \rm (2010):40 
\bibitem {Birks} J.B. Birks, Proc. Phys. Soc. \rm \bf A 64 \rm(1951):874
\bibitem {Cozzini2004} C. Cozzini, {\it et al.}, Phys Rev.  \rm \bf C 70\rm (2004):064606
\bibitem {Ales-97} A. Alessandrello, {\it et al.},Phys Lett  \rm \bf B 408\rm (1997):465
\bibitem {Coron-2008} N. Coron, {\it et al.}, Phys Lett  \rm \bf B 659\rm (2008):113       
\bibitem {Ang-2005} G. Angloher,  {\it et al.},    Astropart. Phys.   \rm \bf  23 \rm (2005):325      
\bibitem {Benoit-2007} A. Benoit, {\it et al.},  Nucl. Instr. and Meth. \rm \bf A 577 \rm(2007):558
\bibitem {Coron-2008-2} N. Coron, {\it et al.}, J. Low Temp. Phys. \rm \bf  151 \rm (2008):865
\bibitem {Gironi09} L. Gironi, C. Arnaboldi, S. Capelli, O. Cremonesi, G. Pessina, S. Pirro and M. Pavan, Opt. Mat. \rm \bf 31 \rm Issue 10 (2009):1388   
\end{thebibliography}
\end{document}